\input{aipcheck}

\documentclass[
    ,final            
  ]
  {aipproc}

\layoutstyle{6x9}

\begin{document}

\title{The CHilean Automatic Supernova sEarch (CHASE)}


\author{G. Pignata}{
  address={Universidad de Chile, Camino El Observatorio 1515, Casilla 36-D Santiago, Chile}
}

\author{J. Maza}{
  address={Universidad de Chile, Camino El Observatorio 1515, Casilla 36-D Santiago, Chile}
}

\author{R. Antezana}{
  address={Universidad de Chile, Camino El Observatorio 1515, Casilla 36-D Santiago, Chile}
}

\author{R. Cartier}{
  address={Universidad de Chile, Camino El Observatorio 1515, Casilla 36-D Santiago, Chile}
}

\author{G. Folatelli}{
  address={Universidad de Chile, Camino El Observatorio 1515, Casilla 36-D Santiago, Chile}
}

\author{F. Forster}{
  address={Universidad de Chile, Camino El Observatorio 1515, Casilla 36-D Santiago, Chile}
}

\author{L. Gonzalez}{
  address={Universidad de Chile, Camino El Observatorio 1515, Casilla 36-D Santiago, Chile}
}

\author{P. Gonzalez}{
  address={Universidad de Chile, Camino El Observatorio 1515, Casilla 36-D Santiago, Chile}
}

\author{M. Hamuy}{
  address={Universidad de Chile, Camino El Observatorio 1515, Casilla 36-D Santiago, Chile}
}

\author{D. Iturra}{
  address={Universidad de Chile, Camino El Observatorio 1515, Casilla 36-D Santiago, Chile}
}

\author{P. Lopez}{
  address={Universidad de Chile, Camino El Observatorio 1515, Casilla 36-D Santiago, Chile}
}

\author{S. Silva}{
  address={Universidad de Chile, Camino El Observatorio 1515, Casilla 36-D Santiago, Chile}
}

\author{B. Conuel}{
  address={Van Vleck Observatory, Wesleyan University, Middletown, CT 06459}
}

\author{A. Crain}{
 address={University of North Carolina at Chapel Hill, Campus Box 3255, Chapel Hill, NC 27599-3255} 
}

\author{D. Foster}{
 address={University of North Carolina at Chapel Hill, Campus Box 3255, Chapel Hill, NC 27599-3255} 
}

\author{K. Ivarsen}{
 address={University of North Carolina at Chapel Hill, Campus Box 3255, Chapel Hill, NC 27599-3255} 
}

\author{A. LaCluyze}{
 address={University of North Carolina at Chapel Hill, Campus Box 3255, Chapel Hill, NC 27599-3255} 
}

\author{M. Nysewander}{
 address={University of North Carolina at Chapel Hill, Campus Box 3255, Chapel Hill, NC 27599-3255} 
}

\author{D. Reichart}{
 address={University of North Carolina at Chapel Hill, Campus Box 3255, Chapel Hill, NC 27599-3255} 
}

\begin{abstract}
The CHASE project started in 2007  with the aim of providing  young southern supernovae (SNe) to the Carnegie Supernova Project (CSP)  and  Millennium Center for Supernova Studies (MCSS) follow-up programs.
So far  CHASE has discovered 33 SNe with an average of more than 2.5 SNe per month in 2008. In addition to the search  we are carrying out a follow-up program targeting bright SNe. Our fully automated  data reduction allows us to follow the evolution on the light curve in real time, triggering further observations if something potentially interesting is detected. 
\end{abstract}

\maketitle


\section{Survey aim}

The CHASE survey started on March 2007 with the goal of  providing  young southern supernovae SNe to the CSP \cite{Hamuy06} and  MCSS follow-up programs.
Those projects are obtaining systematic photometric, spectroscopic and polarimetric observations of nearby SNe to better understand the physics of the explosion, their progenitor, detecting possible sources of biases when using very distant SNe with large look-back times as distance indicators.
 CHASE focuses on southern hemisphere because, historically,  the SNe discovered in the northern hemisphere  largely exceed  those discovered in the southern hemisphere (see Fig.~1).
We also aim to discover SNe as young as possible because it is at these early phases where it is possible to obtain crucial information about the explosion and the progenitor.

 \begin{figure}
  \includegraphics[height=.23\textheight]{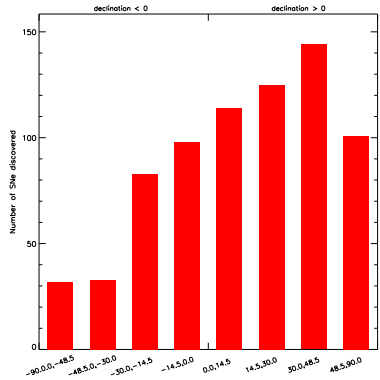}
  \caption{ From 2001 to 2007 68\% of the SNe discovered were found in the northtern hemiphere and 32\% in the southtern hemiphere. In every bin there is the same amount of sky surface, therefore the number of SNe per bin should be constant. The histogram shows clearly the need of a southern SNe search.}
\end{figure}

\section{Instrumentation, data acquisition and reduction}

To carry out the CHASE survey we are using four of the six Panchromatic Robotic Optical Monitoring and Polarimetry Telescopes (PROMPT).
These 40-cm telescopes are 100\% computer controlled and have been designed to get prompt observations of the optical counterparts of  Gamma Ray Burst. These characteristics make the PROMPTs  an ideal tool to carry out a nearby SN search where the instrument has to
move quickly from one galaxy to another. 
The field of view of the PROMPTs is $\sim$ 10'x10' with a pixel size of  $\sim$ 0.6''. The  overhead between two consecutive exposures on the same position is $\sim$  10 seconds.
 Using the images collected so far, we have estimated a limiting magnitude  of $\sim$18.0  with a 40 seconds exposure.\\
Target selection, data acquisition, download and reduction is fully automated. This guarantees that the search is performed in real time, which is necessary to find very young SNe. 
Every day a code selects  a list of galaxies to put in the various telescopes queues, based on the characteristics of each facility, the observations carried out on previous nights, the target visibility and our priority scale.
A couple of hours before twilight our pipeline automatically makes a query to the PROMPT ftp server to download the most recent calibration images (Darks and Flats).
Then, as soon as the scientific images are acquired by the instruments and archived, they are automatically downloaded on our reduction machines at Cerro Cal\'an. Currently the data are processed on a double processor, dual core machine that allow us to fully reduce the data, usually before noon of the following day.
For the data reduction we developed  a modified version of the pipeline used by the ESSENCE survey \cite{Miknaitis07}. 
In short, the data reduction flow consists of:\\
1) dark subtraction and flat field correction, 2) images astrometrization, 3) image registration, 4) image combination, if more than one image per field have been taken, 5) objects detection and zero point calibration, 6) template subtraction, 7) candidates selection, 8) candidates web page generation.

The visual  inspection is performed by the members of the CHASE team and by undergraduate students who have expressed interest in the project. If a good candidate is detected it is added to the corresponding telescope queue for the next night with  maximum priority.  If the candidate is  present in the confirmation image, it is included in a IAU circular and, thanks to the tight collaboration with the CSP, a classification spectrum is acquired using the Las Campanas Observatory facilities. 
Base on the spectroscopic classification, young  SNe are intensively followed up by the MCSS and CSP collaboration (see http://csp1.lco.cl/$\sim$cspuser1/PUB/PROJ/SN\_07\_08.html).\\

\section{Survey strategy and results}

Our relatively shallow limiting magnitude reduce considerably the space volume in which we can detect very young  SNe. The galaxies with radial velocity < 3000 km s$^{-1}$ thus constitute the sample within which we search for these young objects.
Through montecarlo simulations we have estimated that an observational  cadence of 3-4 days is the best strategy for the purpose.
In addition to this golden sample of very nearby galaxies we have a more extended sample which radial velocity is in general  $<$ 8000 km s$^{-1}$. Those galaxies are our backup targets to fill visibility gaps during the night and through the year. They are observed with a cadence which could vary from five days to more than a month.
On average we  observe $\sim$ 250 galaxies per night. However, since our priority in the PROMPT scale is low, this number is highly variable form night to night, significantly decreasing during dark time. Indeed we have little control on which galaxies are observed during a given night and  this do not allow us to fully implement our search strategy. Nevertheless, for every night  we carefully keep truck of what was observed and under which conditions. This, together with a complete characterization of the galaxies in our sample,  will allow us to compute a very well constrained SN rate once a sufficient number of SNe get discovered.\\  
\noindent
After a difficult start on March 2007 when only two of the four PROMPTs were simultaneously  in operation, since October 2007 all four facilities have been operating. On the same month the PROMPT team also managed to considerably improve the telescopes pointing greatly increasing the overlap between images obtained at different epochs. Under these better operational conditions on November 2007 we discovered our first two SNe. On 2008  we have discovered other 31 SNe with an average of more than 2.5 SNe per month.
Those numbers already make CHASE the most successful survey operating in the southern hemisphere. 

Using the classification spectrum to set the phase of our SNe we estimate that among the sixteen Type II SNe discovered, ten where detected within a week after the explosion. Concerning Type I SNe, among the seventeen discovered, eight were detected before maximum light. In particular, the Type Ia SN~2008hv was detected fifteen days before maximum, making this object one of the youngest type Ia ever discovered.   
  
\begin{figure}
  \includegraphics[height=.23\textheight]{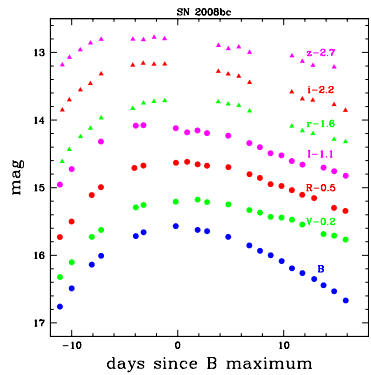}
  \caption{Preliminary BVRIr'i'z' light curves of Type Ia SN~2008bc obtained with the PROMPT telescopes. The SN was discovered  by CHASE 11.7 days before maximum light.}
\end{figure}

\subsection{Follow-up program}

In addition to the SN search,  we are following  bright SNe in the UBVRIu'g'r'i'z' filters (see Fig.~2). Our fully automated  data reduction procedure allow us to follow the evolution of the light curve in real time, triggering further observations if something potentially interesting  is detected. We will also compare the resulting light curves with those produced by the CSP in order to remove any systematic error which may affect the final photometry.

\section{Future plans}

{\bf Optical search:} In spite of the excellent results obtained with the PROMPTs the observations scheduling does not allow us to fully implement our search strategy and therefore completely achieve our goals. 
For this reason around middle 2009 we will enhance the CHASE search and follow-up capabilities with our own 50-cm telescope where we will have full control on the queue organization.\\

\noindent
{\bf Near infrared search:} The University of Tokyo plans to operate an astronomical observatory on the top of  Cerro Chajnantor at  5600m. A 1-m pilot telescope optimized for infrared observations should start to operate  around middle of 2009. The superb infrared sky transparency of this high altitude site makes this facility an ideal tool to carry out a near infrared SN search on nearby star-burst galaxies with the aim to discover extinguished SNe. \\


\begin{theacknowledgments}
G.P. acknowledges support by the Proyecto FONDECYT 3070034. G.P, F.F., M.H., G.F.
and J.M N\'ucleo Milenio P06-045-F funded by Programa
Bicentenario de Ciencia y Tecnolog\'ia from CONICYT and Programa
Iniciativa Cient\'ifica Milenio from MIDEPLAN. M.H and  J.M acknowledges partial support from Centro de Astrof\'\i sica FONDAP 15010003 and from Center of Excellence in Astrophysics and Associated Technologies (PFB 06). G.F acknowledges partial support from Comit\'e Mixto ESO-GOBIERNO DE CHILE
\end{theacknowledgments}



\bibliographystyle{aipproc}   

\bibliography{pignata_1}

\IfFileExists{\jobname.bbl}{}
 {\typeout{}
  \typeout{******************************************}
  \typeout{** Please run "bibtex \jobname" to optain}
  \typeout{** the bibliography and then re-run LaTeX}
  \typeout{** twice to fix the references!}
  \typeout{******************************************}
  \typeout{}
 }

\end{document}